# The Human Factor in AI Safety

**Morteza Saberi**

School of Computer Science, University of Technology, Sydney, NSW, Australia
Morteza.saberi@uts.edu.au

## Abstract

AI-based systems have been used widely across various industries for different decisions ranging from operational decisions to tactical and strategic ones in low- and high-stakes contexts. Gradually the weaknesses and issues of these systems have been publicly reported including, ethical issues, biased decisions, unsafe outcomes, and unfair decisions, to name a few. Research has tended to optimize AI less has focused on its risk and unexpected negative consequences. Acknowledging this serious potential risks and scarcity of research I focus on unsafe outcomes of AI. Specifically, I explore this issue from a Human-AI interaction lens during AI deployment. It will be discussed how the interaction of individuals and AI during its deployment brings new concerns, which need a solid and holistic mitigation plan. It will be discussed that only AI algorithms' safety is not enough to make its operation safe. The AI-based systems' end-users and their decision-making archetypes during collaboration with these systems should be considered during the AI risk management. Using some real-world scenarios, it will be highlighted that decision-making archetypes of users should be considered a design principle in AI-based systems.

## Introduction (Section 1)

AI and its backbone technologies are advancing on an unprecedented scale (Schiffet. al. 2020). This fast development of AI models leads to the emerging of various AI-powered products and services. Top tech companies, including Google, Amazon, and Apple, develop and use these products to provide better services to their customers, gaining significant publicity. This publicity reinforces the role of AI in fast development for other companies including, small-medium enterprises across various sectors (Hansen, & Bøgh 2021).

At the same time, there are serious discussions and debates on AI misconducts or shortcomings such as unfair outcomes, gender discrimination, biased decisions, dangerous actions (Turchin, & Denkenberger 2021). Since this backlash primarily targeted the AI community, various new research directions and domains have emerged as a response including, AI fairness, Explainable AI, Human-AI interaction, Ethical AI, and AI safety (Foulds et al. 2020, Mathews 2019., Amershi et al. 2019). However to have the robust, fair, safe, and responsible AI a collective effort is needed among its stakeholders.

Figure 1 depicts an exemplar AI-powered product's main two stages or lifecycle. The first part is their development stage, which may involve the specific dataset of the company. For simplicity, the team that developed the AI-powered product is named the "service provider," and it has been assumed they are sitting outside the company. The service provider either produces the product on an (i) large scale or develops an (ii) customized one. Referring to those two stages, the main focus of AI safety literature is the "development stage," and there is not much work on the second part, the "deployment stage." Thereby the devised safety mechanism in the development stage is only half of the solution and safety guard.

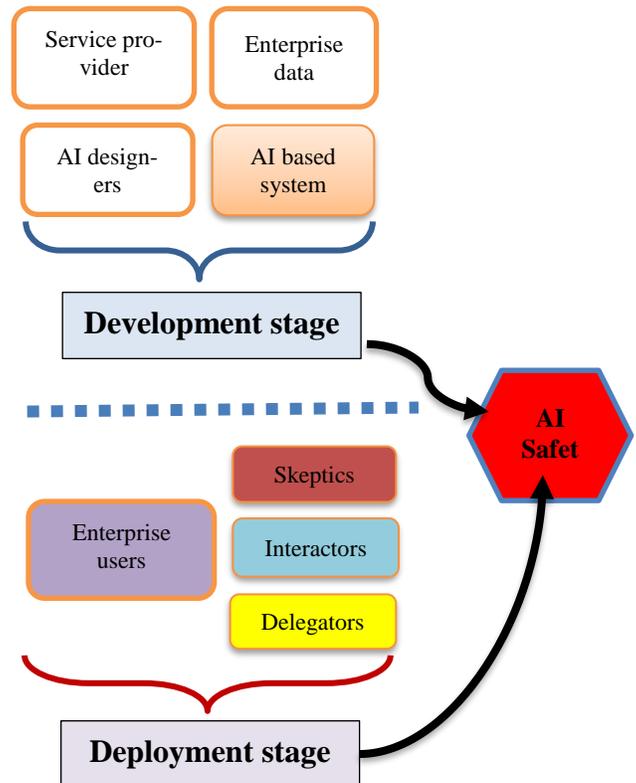

Figure 1. AI-powered product lifecycle

Some of the main AI safety challenges during the "deployment stage" from the human interaction lens will be discussed in this work. This is due to the AI safety literature gap in considering the safety issues related to end-users perspectives, needs, interactions, and roles. AI will be more robust and safe when the AI developers consider its lifecycle challenges. AI-Powered products' end users are naturally diverse, ranging from the organization's top managers, middle managers, knowledge workers to their customers. They also have diverse attitudes towards AI-Powered products and different experiences and expertise. Among the heterogeneous end-users, we have cohorts with (i) No technical background of AI, (ii) general understanding of AI models, (iii) being skeptical about AI and its power, (IV) believing in AI as an ultimate solution, to name a few. Unfortunately, there is not much work, tools, or methods that can support top managers in choosing the right AI-Powered product according to their organization's need and culture. That's why looking at AI as a service enabler that leads to the product or service is essential in those newly emerged fields of AI and AI safety.

Specifically, the impact of knowledge workers with different decision-making archetypes on AI safety will be discussed. To this end, three types of knowledge workers' attitudes towards AI-powered products will be considered as depicted in Figure 1, namely, "Skeptics," "Interactors, "and "Delegators." Section 2 is devoted to describing these decision-making archetypes with two toy examples that highlight their possible threads towards AI safety. In Section 3, classification models will be used as a reference. It will be discussed how changing the classifier metric from "precision" to "recall" makes a given AI-Powered product safer for some cases and riskier for other scenarios. The possible consequences of these risks in not considering the knowledge workers' decision-making archetypes have been presented in Section 4 regarding the confusion matrix. A holistic risk map for each decision-making archetypes will be presented in Section 5 when we have an automatic AI product in place. The paper will be concluded with a short discussion on the mitigating strategy during the "deployment stage," considering the type of involved knowledge workers.

## Three decision-making archetypes in enterprises: From AI Safety Perspective (Section 2)

The AI-based systems assist the organizations in their operations, categorized in two main modes: (i) decision augmentation and (i) automatic decision-making (Leyer, & Schneider 2021). While in the first one, the amount of knowledge workers' involvement and authority is higher than the second mode, the involvement of knowledge workers in the second one is essential. Thus, for both modes, the human element should be considered for any AI safety plan. I am using the taxonomy which Meissner and Keding proposed in their paper by classifying the AI users into three categories, namely, skeptics, interactors, and delegators:

a. "**Skeptics** do not follow the AI-based recommendations."
b. "**Interactors** are open to the use of AI but do not rely on it entirely."
c. "**Delegators** largely transfer their decision-making authority to AI."

Figure 2 depicts how a given knowledge worker with the above decision-making archetypes interacts with an AI-based system.

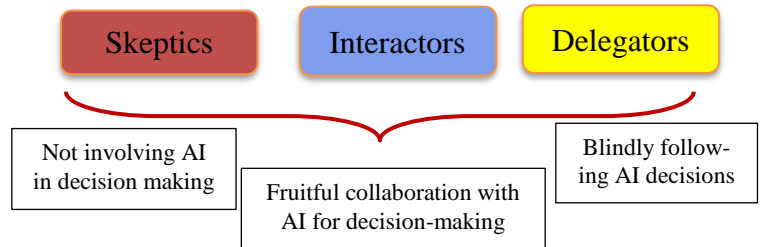

Figure 2. Three decision-making archetypes and their interaction with AI based systems

If the knowledge workers keep ignoring AI decisions, even the safest AI models can be part of accident reasons within a company. On the other hand, if the knowledge workers blindly believe in AI decisions, they may see them as a concrete action plan rather than a recommendation, which needs their final approval. Let's have a simple example of an AI-powered product safety issue by having the involvement of a **skeptic** knowledge worker:

**Example 1.**
"*John is an experienced engineer who is working for a mining company. Recently the company launched an AI-based alert system for the evacuation. John believes most of the system's alert is false and keeps ignoring them. This ultimately led to the injury of some of the mineworkers and put John in the spot*"

The second example shows how a wrong misconception of a given **delegator** knowledge worker about AI power and accuracy leads to an unsafe outcome:

**Example 2.**
"*Jess is a new graduate who is a fan of AI with a general understanding. Jess has recently joined an insurance company. The company uses a customized AI-based system, filtering scam emails. The blind following AI model cost a lot jess in a phishing attack, which led to a severe company financial data leaking, and she had to change her organization at the end.*"

The following section will reinforce the importance of considering the deployment stage during the AI product development to make AI-powered products safe from the Human involvement lens (Meissner, & Keding 2021). It shows how simply changing a classification performance metric to another one makes a given AI-based product safer for a given user who is skeptical/delegator towards an AI-powered product.

## The importance of decision-making archetype consideration in classifier performance measurement (Section 3)

AI-based classification is one of the main tasks of AI models which have been reached to good maturity and accuracy. Two main accuracy metrics look at the accuracy from different angles, namely, "recall" and "precision" (Davis, J., & Goadrich 2006). This section will discuss how consideration of knowledge worker attitude toward AI during the accuracy metric setting may alleviate part of the AI safety risks. Two examples have shown the trade-off between "recall" and "precision" from an AI safety perspective. To this end, a binary classifier has been used as an example, assuming it is operating in the "augmentation mode":

- **Example 3**: The second example has been used as a reference in Example 3, which will figure out for a "**delegator**" personality which accuracy metric is safer.
- **Example 4**: In this example, the same type of user has been considered, a **delegator**. However, due to the new setting of the classifier model, its safe accuracy metric is different from the one in Example 3

**Example 3.**
Assume that Jess's company received 260 emails in a single day, 60 of which are *scam emails*. There are two AI models which aim to flag out the "*scam emails*" with the below confusion matrixes:

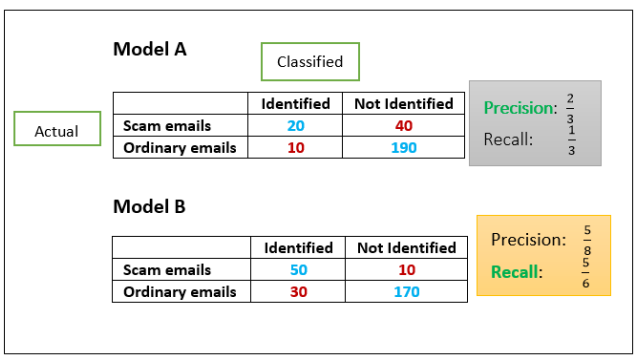

Figure 3. Confusion matrixes of Example 3

Model B is much safer for Jess, **delegator,** considering it has the less False Negative error rate. Since she may blindly follow AI, the scam email the AI system has not flagged out may deceive her easily. To this end, if the "recall" is the basis of model selection, part of the safety issue will be resolved.

**Example 4.**
*Elizabeth is a mother of a nine years old girl and is currently working part-time as a teacher. She is a fan of YouTube primarily because of its powerful movie recommender system. Elizabeth just realized that her girl, Emma, sometimes likes to watch movies with her. Elizabeth sometimes has to check her email and leave Emma alone to watch the rest of the film.*

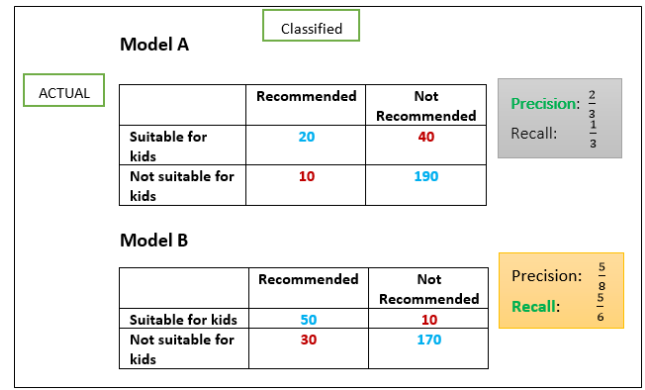

Figure 4. Confusion matrixes of Example 4

Let's see which metric suits Elizabeth's movie recommender system. For simplicity, assume there is a total of 260 free movies available on YouTube which 60 of them are *suitable for kids*. There are two AI models, which identify the "*suitable for kids*" movie with the below confusion matrixes:

Since she may blindly follow AI, she accepts any AI recommender system recommendations. Thus a confident recommender system, higher precision, should be used by Elizabeth. Therefore Model A is much safer for Elizabeth, **delegator.**

**Note**. Per these two examples, It is evident that the preferred performance metric is different for the same decision-making archetype. The reason is that for Example 3, the AI system is picking a high-risk entity, while in the second one, it is recommending a safe entity. Thus, just the structure of the classifier is also of importance in using the correct performance measurement metric.

## How decision-making archetype change the consequence of AI based decision? : Confusion matrix perspective (Section 4)

Now it is clear that organizations should identify the AI-powered product safety issues and proactively mitigate them. To this end, they need a holistic risk assessment plan, which ultimately makes those risks more visible and easier to be dealt with. The risk matrix is a well-known method that has been widely used in academic and industry entities, and its taxonomy will be used in this section (Garvey, & Lansdowne 1998). This assists us in measuring the risk consequence of AI-based system decisions for "**Skeptic**" and "**Delegator**" knowledge workers. Those consequences can be one of the following ones, namely, insignificant, minor, moderate, major, and catastrophic. The confusion matrix shows how different outcomes of a given classifier lead to an unsafe situation. Example 1, Insurance Company, is used for this mapping and presented the risk consequences.

Table 1. The consequence of AI based decision for skeptic and delegator decision-making archetypes

|  | **Skeptic** | **Delegator** |
|---|---|---|
| **False Positive** | NA | The range of **insignificant to major** |
| **True Positive** | The range of **minor** to **catastrophic** | NA |
| **True Negative** | NA | NA |
| **False Negative** | NA | **Moderate to Catastrophic** |

Lets explore some of these consequences:
- **Skeptic Knowledge worker**: Since the Skeptic knowledge worker ignores the AI-based system recommendation, just the consequence of such interaction has been reported in Table 1. If the organization is unaware of such an attitude, while the AI system correctly identifies the scam email, the knowledge worker may expose the organization to this risk. There is a range of consequences out of this attitude:
    - **First scenario/ Minor to moderate**: The Skeptic knowledge worker is referring the scam email to her colleague, either "interactor" or " delegator" and acting per AI system recommendation.
    - **Second scenario/ Major and catastrophic**: The Skeptic knowledge worker responds to the email and exposes the organization to severe risk.
- **Delegator Knowledge worker:** when the knowledge worker blindly accepts AI system recommendations, the AI errors directly expose the organization to the Scam email risks.
    - **False-positive:** The Email is wrongly classified as a "scam email." Thus based on the content and intention of the Email, it may expose the organization to a different range of consequences. For example, if it was related to the payment, ignoring it may lead to some fees penalties, losings its credit, customers disappointment, to name a few. These can be placed as moderate to major consequences. The consequence can also be insignificant or minor when the Delegator Knowledge worker refers it to her interactor Knowledge worker, who again checks the AI system's recommendation.
    - **False-negative:** The consequence of this error generally is higher than the "false positive" error. If the Delegator Knowledge worker responds to this "scam email," the consequence can be from the range of Major to catastrophic based on the harmful level of the Scam email. If the knowledge worker refers to her interactor Knowledge worker, the consequence can be less severe if the former identifies its malicious intention.

The discussed the consequences of AI system safety issues could be severe for the organization when "**Delegator**"/"**Skeptic**" knowledge workers are in place. Thus, those organizations should plan to align their knowledge workers with the AI system before deployment. Otherwise, we should expect a couple of failures even with a robust and accurate AI-based system due to the misalignment, specifically in the high stake context.

**Note**. It should be noted the three types of end-user's attitude towards are the main one while we have some subhierarchy under each of them. For example, the person may show herself be the Interactors while actually she is some between "Interactors" and "Delegators". For example Stevenson and **Doleac studied how judges** are interacting with an algorithmic risk assessments which have been adopted as an aid to judicial discretion in felony sentencing. They found that " judges' decisions are influenced by the risk score, leading to longer sentences for defendants with higher scores and shorter sentences for those with lower scores". Those judges can be categorized as end users which are interacting by AI model but at the same time transfer part of their authority to the algorithmic risk assessment model.

# Automatic decision-making based AI systems: The role of decision-making archetype from AI Safety perspective (Section 5)

In Section 4, the consequences of AI safety have been discussed for the "augmenting mode" having "**Delegator**"/"**Skeptic**" knowledge workers in place. Those consequences will be studied for the "automatic mode" in this section. As a general rule, the safety check should be much more demanding when the AI system is used in the automatic mode. Studying the consequences of these systems safety issues in the automatic mode is the first step for a rigorous safety check. The outcome of such investigation can be a "binary decision": "**changing the operation mode from "automatic to the augmentation mode**." While Automatic AI systems themselves may raise many safety issues and accidents, considering their post-deployment stage may add other types of accidents. The consequences of these accidents for the kinds of attitudes towards AI systems are discussed in this section.

- correct application in the automatic mode considering an individual with "interactor" type.
- The lower part shows the consequences when the AI system is deploying wrongly in the automatic mode.

The consequence of the AI system is automatic when we have knowledge workers with "skeptics" or "delegators" personalities pretty similar since both of them naturally do not interact with the AI system, figures 6 & 7.

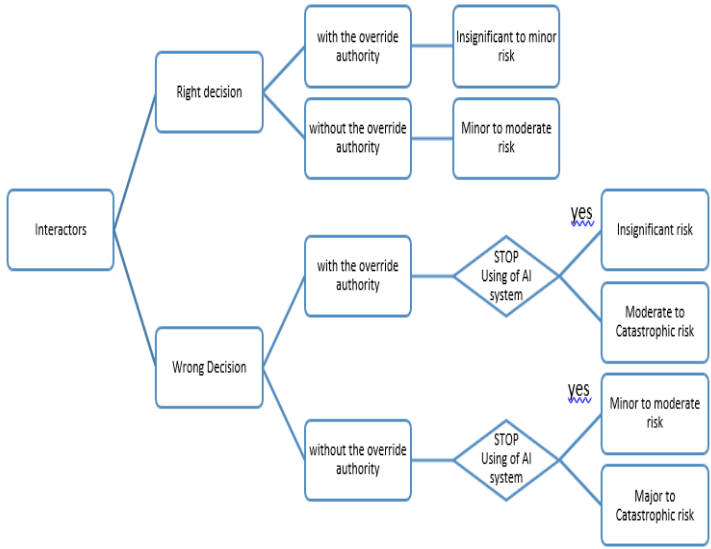

Figure 5. The consequences of AI base system risk's accidents during interaction with "interactors"

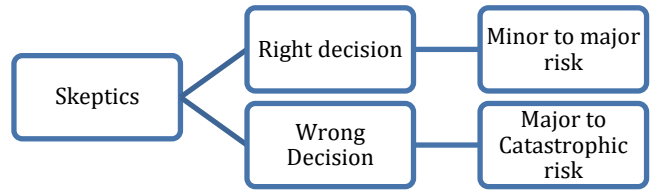

Figure 6. The consequences of AI base system risk's accidents during interaction with "Skeptics"

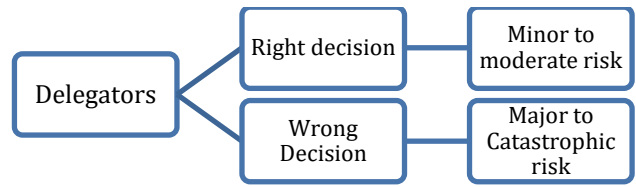

Figure 7. The consequences of AI base system risk's accidents during interaction with "Delegators"

Firstly, the consequences of AI base system risk accidents in the automatic mode during interaction with "interactors" is discussed. Figure 1 depicts the consequences of such deployment under different scenarios. It starts with the assumption that whether using the AI system in such a mode is the "right decision" or not. It is evident that the consequence of AI base system risk accidents, when used in the wrong mode of automation, is much higher than when its operation in the automatic mode is correct. Another factor that changes those consequences is the authority power of the knowledge worker. The interactor knowledge worker with the "override" authority can mitigate some accidents on the spot; otherwise, the consequences of accidents are much severe. If the knowledge worker decides to fight with the decision and stop using AI or prevent its operation, the risk should be much lower compared with the time when the individual cannot stop it. Thus, in summary:

- The upper part of the figure lists the possible consequences of AI system when is deployed for the

**Note**. The listed consequences of Automatic AI base system risk's accidents during interaction with "Delegators", "Interactors," and "Skeptics" are the high-level ones. Each organization should expand such a matrix with a more fine-grained approach by going down to sub-tasks of those AI base systems considering the type of knowledge workers interacting with the AI system sub-tasks.

## Conclusion

By now, it is clear that the human factor itself may make the AI-based decision unsafe, as explored in Section 4 and Section 5. It has been discussed how each decision-making archetype may create a hazardous outcome for two general types of decision making, namely, (i) augmented decision (Section 4) and (ii) automatic decision (Section 5). The organizations can develop a customized risk matrix using the ones presented in Table 1 and figures 6-8 for the specific AI-based system and a given decision. They can identify the knowledge worker who interacts with the system and propose a mitigation strategy based on their decision-making archetype. One effective strategy is revising the job allocation using the matrix risk inputs. This may lead to assigning knowledge workers whose decision-making archetypes align with the AI-based system's operation within the organization. Another strategy is implementing a proactive oversitting mechanism to mitigate detected risks. In some cases, the organization may change its process to be aligned with its human resources and capacity of the AI-based system.